\documentclass[longbibliography,aps,prx,twocolumn]{revtex4-1}

\usepackage{graphicx} 
\usepackage[english]{babel}
\usepackage[T1]{fontenc}
\usepackage[usenames,dvipsnames]{xcolor}
\usepackage{setspace}
\usepackage[compact]{titlesec}
\usepackage{hyperref}
\usepackage{tikz}
\usepackage{amsmath}
\usepackage{amssymb}
\usepackage{mathtools}
\usepackage{appendix}

\newcommand{\fig}{Fig.~\ref}

\newcommand{\eq}{Eq.~\eqref}

\tikzset{
    cross/.pic = {
    \draw[rotate = 45] (-#1,0) -- (#1,0);
    \draw[rotate = 45] (0,-#1) -- (0, #1);
    }
}

\tikzset{
    squar/.pic = {
    \draw[rotate=45] (-#1,0) -- (0,#1) -- (#1,0) -- (0,-#1) -- cycle;
    }
}

\tikzset{
    triag/.pic = {
    \draw (0,#1) -- (0.866*#1,-#1/2) -- (-0.866*#1,-#1/2) -- cycle;
    }
}

\newcommand{\tikzbtriagU}[2][red,fill=red]{\tikz[baseline=-0.5ex]{\path (0,0) pic[line width=1.2pt] {triag=4pt};}}
\newcommand{\tikzbring}[2][red,fill=red]{\tikz[baseline=-0.5ex]{\draw[radius=4pt,fill=black] (0,0) circle; \draw[radius=2.8pt,fill=white] (0,0) circle ;}}
\newcommand{\tikzbtriagD}[2][red,fill=red]{\tikz[baseline=-0.5ex]{\path (0,0) pic[line width=1.2pt,rotate=180] {triag=4pt};}}
\newcommand{\tikzbplus}[2][red,fill=red]{\tikz[baseline=-0.5ex]\path (0,0) pic[line width=1.2pt, rotate = 45] {cross=4pt};}
\newcommand{\tikzbsquar}[2][red,fill=red]{\tikz[baseline=-0.5ex]{\path (0,0) pic[line width=1.2pt,rotate=45] {squar=4pt}}}
\newcommand{\tikzbcross}[2][red,fill=red]{\tikz[baseline=-0.5ex]\path (0,0) pic[line width=1.2pt] {cross=4pt};}

\begin{document}

\title{Curvature-Controlled Geometrical Lensing Behavior in Self-Propelled Colloidal Particle Systems}%

\author{Philipp W. A. Sch\"onh\"ofer}%
\email{pschoenh@umich.edu}
\affiliation{Department of Chemical Engineering, University of Michigan, Ann Arbor, Michigan 48109, USA.}
\author{Sharon C. Glotzer}%
\email{sglotzer@umich.edu}
\affiliation{Department of Chemical Engineering, University of Michigan, Ann Arbor, Michigan 48109, USA.}
\affiliation{Biointerfaces Institute, University of Michigan, Ann Arbor, Michigan 48109, USA.}

\begin{abstract}
\normalsize
In many biological systems, the curvature of the surfaces cells live on influence their collective properties. Curvature should likewise influence the behavior of active colloidal particles. We show using molecular simulation of self-propelled active particles on surfaces of Gaussian curvature (both positive and negative) that curvature sign and magnitude can alter the system's collective behavior. Curvature acts as a geometrical lens and shifts the critical density of motility-induced phase separation (MIPS) to lower values for positive curvature and higher values for negative curvature, which we explain theoretically by the nature of parallel lines in spherical and hyperbolic space. Curvature also fluidizes dense MIPS clusters due to the emergence of defect patterns disrupting the crystalline order inside the clusters. Using our findings, we engineer three confining surfaces that strategically combine regions of different curvature to produce a host of novel dynamic phases, including cyclic MIPS on sphercylinders, wave-like MIPS on spherocones, and cluster fluctuations on metaballs. 
\end{abstract}

\maketitle

\section{Introduction.} 
Mimicry of biological systems has long been a goal in colloidal science. Apart from self-assembly, recreating the dynamics of collective systems -- like the swarming of animals \cite{HH2012, PMRT2014, IU2021}, the collaborative swimming of bacteria \cite{SK2009, BIGKHBA2020} or the dynamics of cell tissue \cite{PJJ2015} -- has received great interest in the last decades. A variety of techniques have been developed to synthesize active colloids \cite{GCBV2008,WM2013, BP2014,DL2019,ADRRFC2021} that harness internal energy and convert it into a self-propelling force, driving the system perpetually far from equilibrium and into novel phases reminiscent of the behavior of their biological counterparts \cite{ WL2012, MBCRD2017,BT2019}. For instance, self-propelled colloidal particles (SPPs) interacting only via excluded volume exhibit motility-induced phase separation (MIPS) into dense but mobile crystalline clusters coexisting with a low-density gas above a critical packing fraction \cite{CT2015, MFHPY2016}. MIPS has been studied predominantly in three \cite{TW2021} or (flat) two dimensions \cite{CT2015}. However, many active organisms live on complex surfaces whose curvature that can affect the collective dynamics. In biology, this relationship between activity and surface curvature is illustrated by the collective flow of embryonic tissue during gastrulation \cite{K2005} or the migration of epithelial tissue in the gut \cite{KEMGRLBHV2019}. Although first steps towards studying active colloids on curved surfaces using computation have been taken \cite{SH2015, AS2017, AZL2019, HEKSS2020} and algorithms have been suggested \cite{PK2016, AS2018, YL2019}, the effect of surface curvature on MIPS has been little investigated. Most studies focus either on polar particles \cite{VCB-JCS1995, SBM2017} or on very low or very high densities \cite{JKL2017, BG2017, PVWL2018}. Additionally, most studies are limited primarily to spherical confinement, aside from a few exceptions \cite{AZL2019, HEKSS2020}, and thus to systems with predominantly positive Gaussian curvature $K>0$. The influence of negative (saddle-like) curvature $K<0$ is unexplored even though such structures are ubiquitous in nature \cite{LTL2003, SSWB2007, ALKD2009, WLS2009, HC2018, SNSDP2021}.

In anticipation of the importance of curvature in synthetic active matter, we investigate the dynamic behavior of hard, non-polar SPPs on both positively and negatively curved surfaces. We performed molecular dynamics (MD) simulations of SPPs in spherical (positive Gaussian curvature) and hyperbolic (negative Gaussian curvature) space by confining them to sphere and gyroid minimal surfaces, respectively. While a sphere has constant positive Gaussian curvature, a surface of constant negative Gaussian curvature cannot be embedded in 3D Euclidean space without self-intersections or singularities. The gyroid, which has negative curvature everywhere, is the closest approximation of the hyperbolic plane in Euclidean space \cite{S-TRCH2003, S-TFH2006}. We first investigate whether curvature shifts the critical density for the onset of MIPS, an important quantity in active matter studies, for low, intermediate and high curvature sphere and gyroid surfaces. We report an interesting geometrical lensing effect for surfaces of intermediate curvature, which we show theoretically arises from the nature of parallel lines in spherical and hyperbolic space. We next show that curvature destabilizes solid MIPS clusters, breaking them up into a dense fluid domain, which we relate to lattice defects dictated by the surface geometry.  Finally, we synthesize our findings to give three examples of how simple, non-spherical confining surfaces may be designed to obtain novel MIPS phenomena. Specifically, we engineer \textit{cyclic motility-induced phase separation} (CMIPS) on spherocylinders, where clusters of SPPs alternately assemble and disassemble (see Movie 1); \textit{cluster waves} on "spherocones", where SPPs alternately assemble at one end and disassemble at the other (see Movie 2); and \textit{cluster fluctuations} on metaballs where large, spatiotemporal fluctuations in cluster size and position occur (see Movie 3). \\

\section{Results}
Throughout the paper we use SPPs with diameter $\sigma$ and persistence length $l=100\sigma$; see definition and simulation methods in the Appendix. As shown in Fig. S1, all observed collective behavior is independent of $l$ for $l>50\sigma$.\\

\subsection{Gaussian curvature shifts the onset of phase separation.}
\label{sec:PhaseSeparation}
\begin{figure*}[t!]
\centering
\includegraphics[width=1\textwidth]{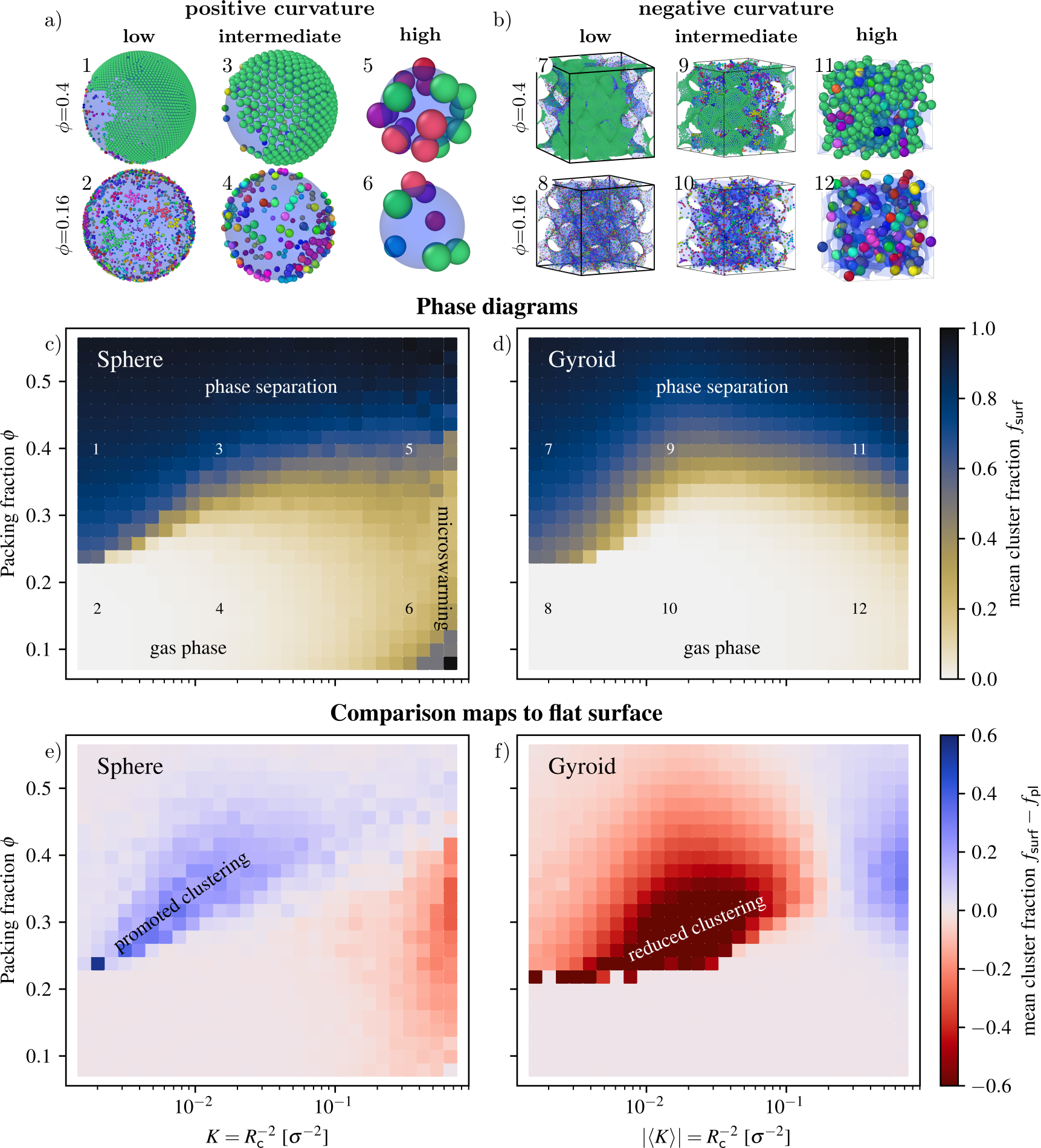}
 \caption{a+b): Snapshots from simulations of the sphere (a) and gyroid (b) systems in the high ($|K|{=}0.34\frac{1}{\sigma^2}$) medium ($|K|{=}0.015\frac{1}{\sigma^2}$) and low ($|K|{=}0.002\frac{1}{\sigma^2}$), curvature regime at two densities $\phi{=}0.16$ and $\phi{=}0.4$. The particles are colored based on their cluster affiliation.  For the cluster construction, two SPPs, $i$ and $j$, are considered neighbors if they interact. The numbers indicate the position within the phase space of (c) and (d). c+d) Color map of the mean cluster fraction $f=\frac{1}{N}\sum_i^N\frac{m_i}{N}$ for SPPs on a sphere surface (c) or on a $2\times 2\times 2$ periodic gyroid (d) with the number of total particles $N$ and the number of members $m_i$ within the cluster of the $i$-th particle. The phase space is sampled based on the packing fraction $\phi$ and the mean curvature $K$. e+f): Comparison color map between the weighted cluster fraction $\Delta f = f_\text{surf}-f_\text{pl}$ of (c)/(d) and of a corresponding system where the SPPs are confined to a flat plane with equal surface area $A_\text{pl}\coloneqq A_\text{surf}$ as the sphere (e) and gyroid (f) surface, respectively. The comparison maps allow us to distinguish between finite-size and curvature effects. Here, blue/red areas indicate curvature regions with higher/lower mean cluster fraction than on their flat counterpart.}
  \label{fig:PhaseDiagram}
\end{figure*}

To study MIPS in relation to the Gaussian curvature of the confining surface, we investigated the collective behavior of SPPs on spheres and gyroids with different mean curvature radii $R_c=\frac{1}{\sqrt{|K|}}$. Our findings are summarized in a phase diagram based on the observed clustering behavior (see \fig{fig:PhaseDiagram} also for clustering definition). The phase diagrams were generated by gridding up the design space linearly within the density range $\phi = [0.08,0.56]$ and logarithmically within the curvature range $|K| = [1.61{\cdot}10^{-3}\sigma^{-2},0.672\sigma^{-2}]$ into a total of 700 points. As shown in the phase diagram, particle clustering directly depends on whether SPPs are confined to surfaces of low curvature ($2\pi|K|^{-\frac{1}{2}} >l$), high curvature ($|K|^{-\frac{1}{2}} < 8\sigma$) or intermediate curvature. SPPs on low-curvature spheres and gyroids behave as though they are confined to a plane as all studied surfaces are locally approximately flat. The behavior of SPPs on surfaces of high and intermediate curvature is more interesting.\\[-0.3cm]

On highly curved surfaces, the collective behavior of the SPP systems is dominated by topological effects. These topological effects are different on the sphere than on the gyroid. On the sphere, MIPS is ill-defined because It is hardly possible to distinguish between fluctuations and clustering due to the small number of particles that can occupy the surface (see also \fig{fig:PhaseDiagram}a). However, the high degree of spatial confinement leads to a spontaneous alignment of SPPs; this microswarming phenomenon was reported in Ref.~\cite{BG2017}. On the gyroid, however, the particles form amorphous networks with only 2-3 contacts per particle (see Fig.~S2 and \fig{fig:PhaseDiagram}b), a markedly different structure than the close packed hexagonal configurations observed in flat space. With hyperbolic (negative curvature) surfaces, the onset density for clustering is higher than in flat space because of the openness of the network-like domains.

Intermediate between the low and high curvature regime, we identify a region where topological effects can be mostly neglected, yet the systems respond to the local confining geometry in opposite ways for spherical and hyperbolic environments. Spherical confinement promotes phase separation and clustering at lower packing densities (see blue region in \fig{fig:PhaseDiagram}e) than in flat space, while the gyroid surface suppresses clustering and the dense phase forms at higher densities (see red region in \fig{fig:PhaseDiagram}f). This contrary effect is explained by the characteristics of spherical and hyperbolic geometry. We designed our simulations such that freely moving SPPs follow geodesic trajectories. In spherical geometry, geodesics correspond to great circles on the bounding sphere where parallel lines do not exist. The pathways of two particles that move in the same direction thus converge and eventually intersect. Consequently, positive curvature acts as a convex lens and increases the probability for SPPs to collide and cluster. In contrast, geodesic trajectories of SPPs traveling in the same direction on a hyperbolic surface are hyper-parallel and diverge from one another. Therefore, negative curvature acts as a concave lens to disperse the particles and suppress clustering.

We use this insight to theoretically predict the critical onset densities of clustering $\phi_c(K)$ based on the growth rate of seed clusters, which are precursors of phase separation. We use a minimal model similar to Ref.~\cite{RHB2013}. Consider a circular seed of radius $r$ within a homogenous gas of SPPs with velocity vector $v_0\hat{\mathbf{v}}$. For seeds to grow into large clusters, the influx of SPPs from the gas phase must exceed the outflux of seed particles back into the gas. We express the evaporation rate by $k_\text{out}(K) = \kappa_\text{out} C(K)$ in terms of the seed circumference $C(K)$ since only particles at the gas-seed interface can leave the cluster. The factor $\kappa_\text{out}$ denotes the average loss rate of particles per unit length. We identify $k_\text{out}$ in Euclidean, spherical and hyperbolic space as\\
      \begin{equation}
      \label{eq:kout}
    \begin{aligned}
    k_\text{out}^\text{Eucl} &= 2\pi r\kappa_\text{out},\\
    k_\text{out}^\text{Sph}(K)  &= 2\pi K^{-\frac{1}{2}} \sin\left(r \sqrt{K}\right) \kappa_\text{out},\\
    k_\text{out}^\text{Hyp}(K)  &= 2\pi |K|^{-\frac{1}{2}} \sinh\left(r \sqrt{|K|}\right) \kappa_\text{out}.\\
    \end{aligned}    
    \end{equation}
\noindent To derive the equation for the absorption rate $k_\text{in}(K)$, we must determine the number of particles that collide with the seed per time unit $\tau$, within which an unimpeded SPP travels roughly a distance $v_0\tau$. Hence, only particles within the shell $r+\sigma \leq x \leq r+\sigma+v_0\tau$ must be taken into account. Since the direction of $\hat{\mathbf{v}}$ is uniformly distributed due to diffusion, a particle at a distance $x$ from the center of the seed is on a collision course with the cluster with probability (using the law of cosines in Euclidean, spherical and hyperbolic trigonometry)\\
      \begin{equation}
      \label{eq:collisionProb}
    \begin{aligned}
     p^\text{Eucl}(x) &= \frac{\arccos\left(\frac{x^2+(v_0\tau)^2-(r+\sigma)^2}{2x(v_0\tau)}\right)}{\pi}\\
     p^\text{Sph}(x) &= \frac{\arccos\left(\frac{\cos((r+\sigma)\sqrt{K}) - \cos(x\sqrt{K})\cdot\cos(v_0\tau\sqrt{K})}{\sin(x\sqrt{K})\cdot\sin(v_0\tau\sqrt{K})}\right)}{\pi}\\
     p^\text{Hyp}(x) &= \frac{\arccos\left(\frac{\cosh(x\sqrt{|K|})\cdot\cosh(v_0\tau\sqrt{|K|}) - \cosh((r+\sigma)\sqrt{|K|})}{\sinh(x\sqrt{|K|})\cdot\sinh(v_0\tau\sqrt{|K|})}\right)}{\pi},\\
    \end{aligned}    
    \end{equation}
    
 \makeatletter\onecolumngrid@push\makeatother
    \clearpage
  \makeatletter\onecolumngrid@pop\makeatother
  
\begin{figure}[t]
\centering
  \includegraphics[width=0.48\textwidth]{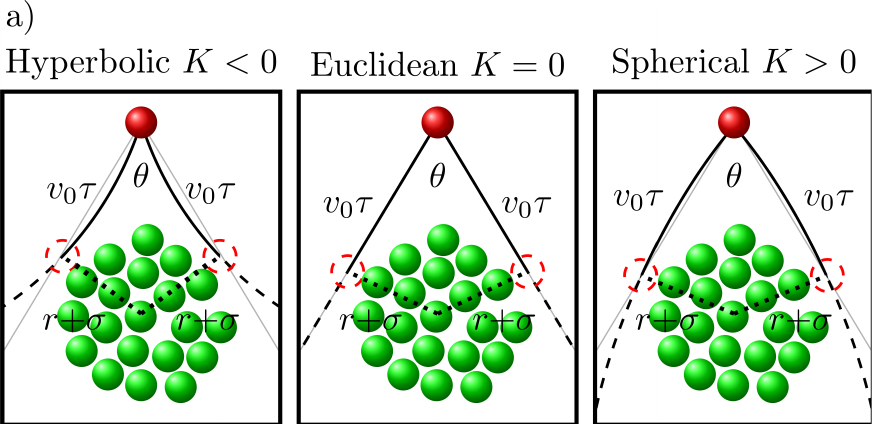}\\[0.2cm]
  \includegraphics{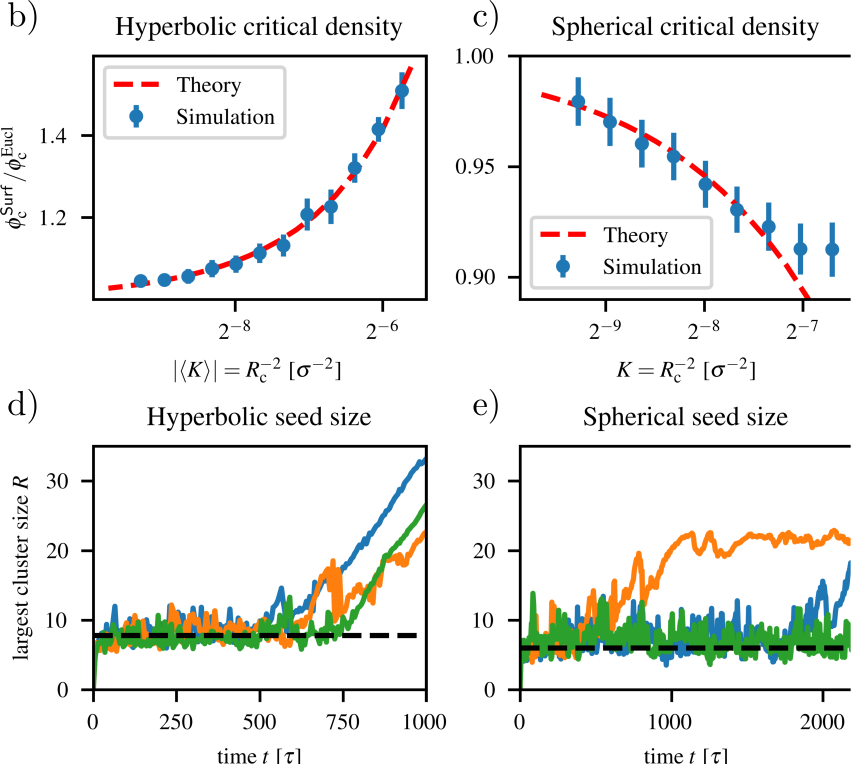}
  \caption{a) Schematic of the collision window of an SPP (red) with a circular seed cluster (green) in two-dimensional hyperbolic, Euclidean and spherical space from left to right. The collision angle $\theta$ of a particle at position $x$ and velocity $v_0$ is derived from the maximal distance a SPP can travel per time $\tau$ and the radius of the cluster $r$. By increasing or decreasing Gaussian curvature $|K|$ the particle trajectories bend towards and away from the cluster, respectively. We use $\theta$ to calculate the collision probabilities in \eq{eq:collisionProb}. b+c) The critical density for phase separation at different curvature $K$ in the medium regime. The onset of clustering is measured on the gyroid (b) and the sphere (c) in relation to flat space and compared to the theoretical predictions of \eq{eq:Prediction} with seed size $r=7.8\sigma$ (b) and $r=6.0\sigma$ (c). d+e) Radius of the largest cluster during different simulations in flat space at the critical density. We approximate the radius as $\sqrt{2}r_\text{g}$ with the radius of gyration $r_\text{g}$. In plot (d) the simulation box size is set to $L=282\sigma$ without noticeable finite-size effects. The seed size before clustering (dashed line) coincides with the choice of $r$ in b). In plot (e) we run the simulation at lower box side length $L=70\sigma$ where finite-size effects are present. The seed size reduces and is in accordance with the choice of $r$ in c).}
  \label{fig:Prediction}
\end{figure}

\noindent See \fig{fig:Prediction}a for a pictorial description of these equations. By integrating over the shell around the seed in the respective spaces and approximating that the gas density is roughly the same as the global density $\phi_g\approx \phi$, we can write:\\[-0.4cm]
      \begin{equation}
      \label{eq:kin}
    \begin{aligned}
    k_\text{in}^\text{Eucl} &= \phi\int_0^{2\pi}\int_{r+\sigma}^{r+\sigma+v_0\cdot\tau} p^\text{Eucl}(x)\cdot x \text{d}x = \phi\cdot P^\text{Eucl}\\
    k_\text{in}^\text{Sph}(K) &= \phi\int_0^{2\pi}\int_{r+\sigma}^{r+\sigma+v_0\cdot\tau}  p^\text{Sph}(x) \cdot \frac{\sin(x\sqrt{K})}{\sqrt{K}} \text{d}x\\
    &= \phi\cdot P^\text{Sph}(K)\\
    k_\text{in}^\text{Hyp}(K) &= \phi\int_0^{2\pi}\int_{r+\sigma}^{r+\sigma+v_0\cdot\tau}  p^\text{Hyp}(x) \cdot \frac{\sinh(x\sqrt{|K|})}{\sqrt{|K|}} \text{d}x\\[-0.1cm]
    &= \phi\cdot P^\text{Hyp}(K).
    \end{aligned}
    \end{equation}
If we assume that $\kappa_\text{out}$ is equal in all three cases and if $k_\text{in}=k_\text{out}$ is in balance at the critical density $\phi_c$, we can predict the shift of $\phi^\text{Sph/Hyp}_c$ on the sphere and the gyroid in relation to $\phi_c^\text{Eucl}$ in Euclidean space. By combining \eq{eq:kout} and \eq{eq:kin}, we get:\\[-0.4cm]
      \begin{equation}
      \label{eq:Prediction}
    \begin{aligned}
    \phi_c^\text{Sph}(K) &= \frac{k_\text{out}^\text{Sph}(K)}{k_\text{out}^\text{Eucl}} \frac{P^\text{Eucl}}{P^\text{Sph}(K)} \phi_c^\text{Eucl}\\
     \phi_c^\text{Hyp}(K) &= \frac{k_\text{out}^\text{Hyp}(K)}{k_\text{out}^\text{Eucl}} \frac{P^\text{Eucl}}{P^\text{Hyp}(K)} \phi_c^\text{Eucl}.\\
    \end{aligned}
    \end{equation}
In \fig{fig:Prediction}(b-c) we fit both functions to our computational results and obtain good agreement between theory and simulation for the seed size $r=7.8\sigma$ on the gyroid and $r=6.0\sigma$ on the sphere. Furthermore, both values coincide with the typical sizes of precursor clusters observed in simulations in flat 2D space at the critical density and before the system starts to phase separate (see \fig{fig:Prediction}(d-e)). We explain the difference in the seed radii by the inherent finite size of the sphere system. The predictions break down at $K^{-\frac{1}{2}}\approx 8\sigma$, where we enter the high curvature regime and can no longer neglect the aforementioned topological effects that act contrary to and eventually even dominate the geometrical lensing.
\begin{figure*}[t]
\centering
  \includegraphics[width=\textwidth]{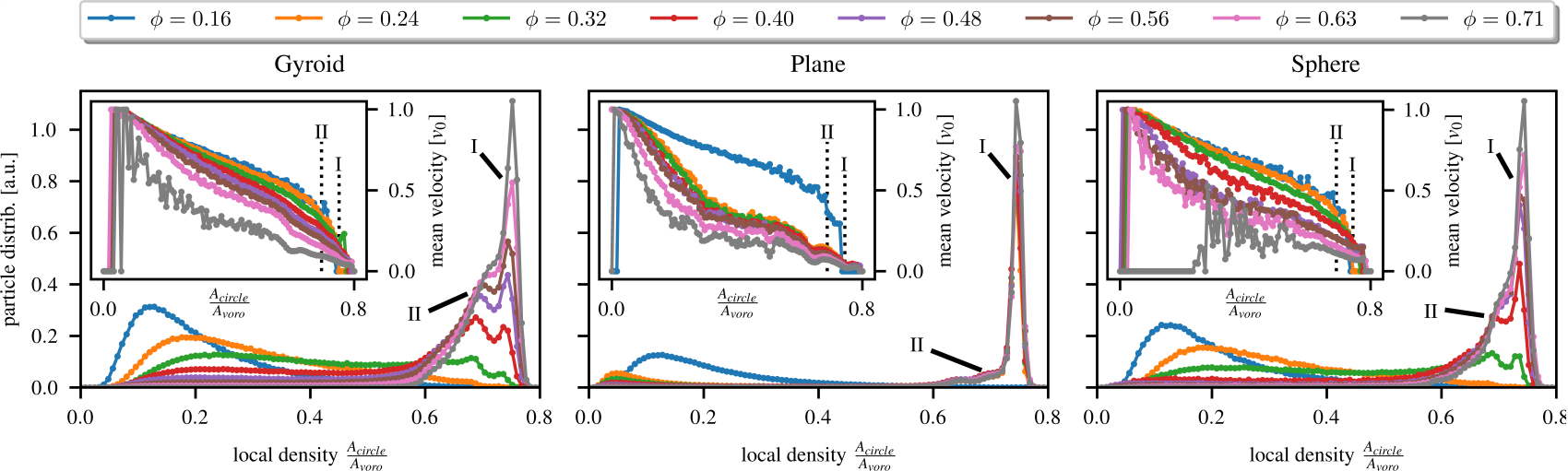}
  \caption{Local density distribution based on the Voronoi diagrams on the gyroid (left), plane (center) and sphere surface (right) at different packing fractions. The peaks annotated with I and II indicate particles within a hexagonal environment and within defect patterns, respectively. The inset plots are the same local density distributions weighted by the mean translation per time step $\tau$.}
  \label{fig:VoroLocalDensity}
\end{figure*}

\subsection{Curvature induces geometric fluidization of dense SPP clusters}
\label{sec:Fluidisation}
In addition to the observed shifted phase separation, curvature also affects the dynamics within the cluster. In flat space, SPPs in the dense phase move within the hexagonally ordered lattice structure via dislocation diffusion \cite{RHB2013}, which follows L\'evy-flight-like dynamics (see SI Fig.~3). However, clusters on both positively and negatively curved surfaces are more mobile than in flat space due to topologically mandated defect scars, which are known to also emerge in packings of hard particles on curved surfaces \cite{BMWSA2015, GKHC2018}. Whereas particles inside a cluster still travel via L\'evy-flight-like transport at low curvature, hexagonal order becomes less prominent by curving the underlying surface further and the system separates not into a gas and a solid phase but instead into a gas and a high density diffusive fluid phase of flowing particles (see SI Fig.~3). We call this mechanism, which is similar to grain boundary melting \cite{S1982, C1983, DBMK2015}, \textit{geometric fluidization} of the cluster. This fluid nature is present even in the active crystal limit where the dense phase spans the sphere or the gyroid.

To directly relate defects to particle motility, we obtained the local density distribution based on Voronoi tessellation in flat and curved systems with $R_c=8.2\sigma$. 
In \fig{fig:VoroLocalDensity} we use the inverse area of the Voronoi cells as an approximation of local density at different packing fractions. In the flat system the local density in the phase separated state is bimodal, which we attribute to the gas and the solid phase. In addition, the distributions on both the gyroid and the sphere indicate a bifurcated high density peak with two sub-domains. Peak I in \fig{fig:VoroLocalDensity} represents tightly packed particles within a mostly hexagonal environment, while peak II, which is dominant at the onset of phase separation, arises from the defect patterns. The position of peak II is consistent with the density distribution in flat space, where, even though no secondary high-density peak occurs, the presence of defects is suggested by a small bump. The second peak in the hyperbolic and spherical systems shrinks with density and, eventually, is indistinguishable from peak I. The disappearance of the defect peak coincides with the kinetic arrest of the active crystal at very high densities. The importance of defect patterns for the fluidization of the clusters becomes even more apparent by relating local density with velocity. The motility of particles decreases with the density of their environments. Closely packed particles are practically arrested within the crystal besides the collective motion of the cluster. At local densities close to peak II we can observe a large increase in velocity in all three systems -- sphere, gyroid and flat space. Therefore, by introducing defects and destabilizing the balance of forces, curvature induces flow and eventually fluidizes the whole cluster.

\subsection{Novel, curvature-induced non-equilibrium phases can be designed and engineered}
\label{sec:Designing}
\begin{figure*}[t!]
\centering
  \includegraphics{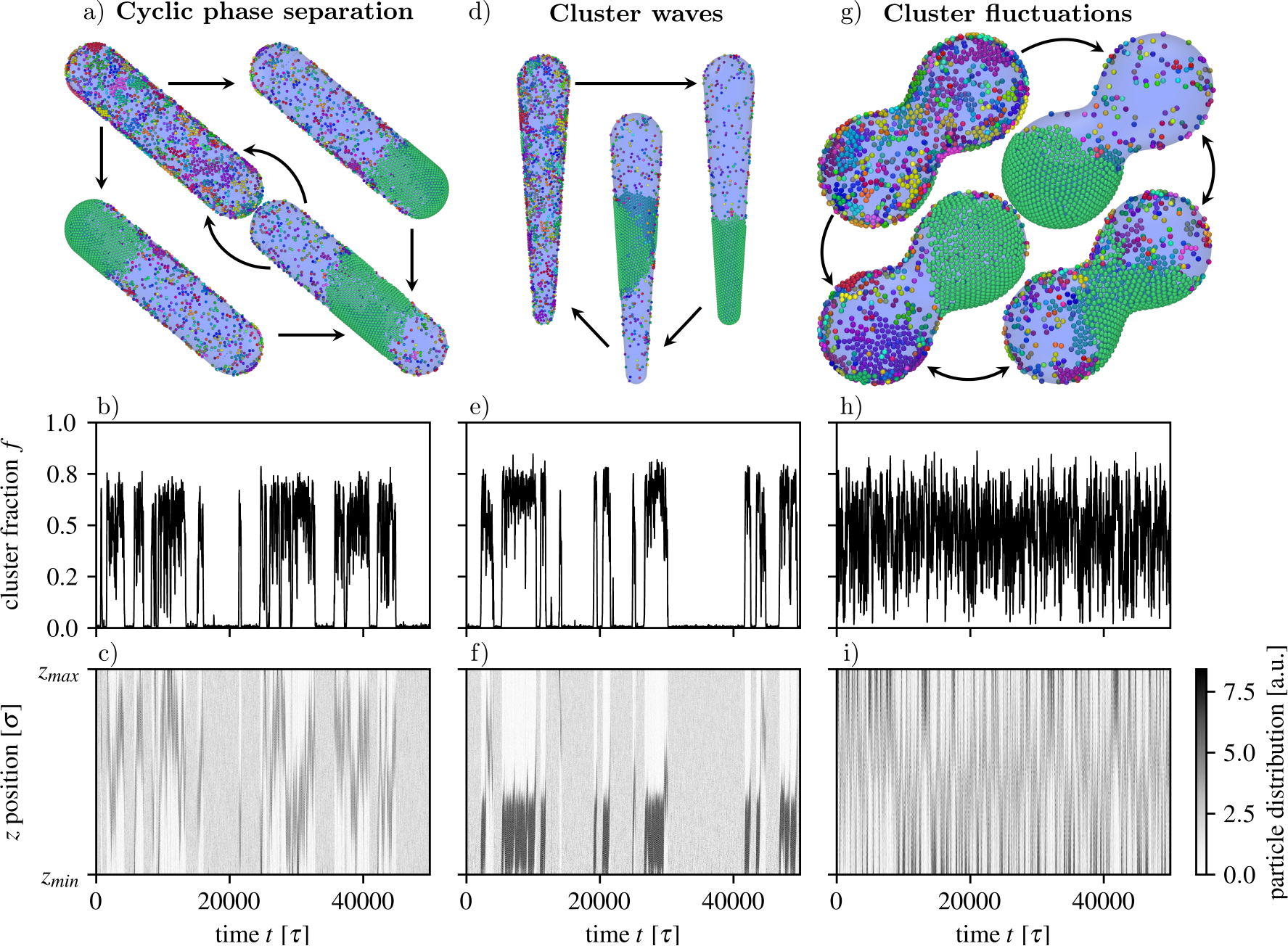}
  \caption{The cluster fraction (center) and the local density (bottom) along the symmetry axis $z$ during a representative simulation of active particles confined to a spherocylider (see Movie 1), spherocone (see Movie 2) and metaball (see Movie 3). The plots indicate different types of non-equilibrium phases as cyclic phase separation (spherocylider), cluster waves (spherocone) and cluster fluctuations (metaball). Schematics of the three different phases are shown at the top.}
  \label{fig:CyclicPhaseSeparation}
\end{figure*}

Based on the above findings, we hypothesize that we can obtain new dynamic phases in SPP systems confined to closed surfaces that strategically combine regions of positive, negative and zero Gaussian curvature. Here we present three such examples, depicted in \fig{fig:CyclicPhaseSeparation} 

\noindent \textit{Cyclic motility-induced phase separation} -- Because cluster stability is different on surfaces of different curvature, we hypothesized that by combining positive and zero Gaussian curvatures in a closed surface we could induce cyclic motility-induced phase separation (CMIPS). A spherocylinder is one such surface, combining regions of positive Gaussian curvature (the spherical caps) and zero Gaussian curvature (the cylinder). We confined SPPs to a spherocylinder of length $l=2\pi R$ and radius $R=10\sigma$. The radius was chosen to produce enhanced phase separation at the caps with $K=0.01\sigma^{-2}$ according to \fig{fig:PhaseDiagram}. As expected, we observe MIPS above $\phi>0.25$, but the SPPs form clusters already at lower densities. In a narrow window  $0.235<\phi<0.25$, the particle density is high enough to induce clustering at the positively curved caps, but low enough to both prevent clustering and destabilize dense regions on the cylindrical part where $K=0$. The different cluster stabilities on different areas of the spherocylinder surface leads to a temporally cyclic phase-separated state as indicated by the cluster fraction and the particle density along the symmetry axis in \fig{fig:CyclicPhaseSeparation}. From an isotropic phase, a cluster starts to grow at one of the spherical caps. This cluster moves away from the caps to the cylindrical part of the surface, where it is not stable and dissolves into the surrounding gas-like phase. These additional particles increase the density of the gas phase such that clusters can reform at the caps (see also \fig{fig:CyclicPhaseSeparation} and Movie 1), and the cycle repeats. Spectral analysis of the time-resolved cluster fraction indicates that the signal of CMIPS can be compared to red noise (see SI Fig.~4). This indicates that the switch between phase separated and gas phase follows a random two state Brownian process \cite{HJ1995}.  

CMIPS is a Brownian-like process as the signal can be compared to red noise (see SI Fig.~4). We observe CMIPS for spherocylinders with radius $R\leq 12\sigma$ ($K\geq6.9\cdot10^{-3}\sigma^{-2}$). For larger radii, the cluster still develops on one of the end caps but no longer dissolves in the zero Gaussian curvature regions. 
 
\noindent \textit{Cluster waves} -- CMIPS can be engineered to create wave-like behavior where the cyclic phase separation is directed always from one side. By decreasing the radius $r$ of one of the caps while keeping the other radius $R$ constant, a second, differently positively curved area on the surface is introduced. On this "spherocone" surface we again observe CMIPS, but the curvature asymmetry biases cluster formation towards the smaller cap as indicated in \fig{fig:CyclicPhaseSeparation} and Movie 2 with $R=10\sigma$ and $r=4\sigma$. After forming at the small cap ($K=0.0625\sigma^{-2}$), the cluster travels towards the large cap into the flat region, dissolves and reforms at the small cap again. We call this phenomenon a ``cluster wave'' as it is reminiscent of the three stages of an ocean wave (build-up, travel, break).

\noindent \textit{Cluster fluctuations} -- Combining surfaces of positive and negative Gaussian curvature provides another strategy for obtaining novel phases. Such a combination can be found in a metaball.  Metaballs are n-dimensional surfaces that meld together into single, contiguous objects. Metaball behavior is seen in cell mitosis, where chromosomes generate identical copies through cell division. Here we confine SPPs to a metaball surface comprising two spheres of radius $R=9.5\sigma$ separated by a distance $l=24\sigma$ such that the two spheres overlap to create a negatively curved "bridge" between them. On the metaball surface we also detect CMIPS below a packing fraction of $\phi =0.28$. Above this density, dense clusters of SPPs form predominantly on the positively curved spherical portions ($K=0.011\sigma^{-2}$) and move into the negatively curved bridge connecting the two metaballs. In this hyperbolic region clusters shrink in size without dissolving completely. From the bridge the cluster then travels either back to one of the two spherical regions, where it grows again. This back and forth between growing and shrinking of the cluster results in high spatiotemporal variations of the cluster fraction, as shown in in \fig{fig:CyclicPhaseSeparation} and Movie 2. Spectral analysis suggests that the fluctuations of the cluster fraction exhibit a frequency scaling behavior reminiscent of pink rather than red noise (see SI Fig. 4). Pink or flickering noise is typical for bistable stochastic processes \cite{HJ1995,EK2009} and ubiquitous in biological and physical systems \cite{SVS2001}.\\

\section{Conclusion}
\label{sec:Conclusion}
In this paper, we studied the effect of negative and positive Gaussian curvature on the motility-induced phase separation of active self-propelled particles. We first introduced curvature by confining the particles to a sphere or gyroid minimal surface. In both cases we identified topological and geometrical effects that alter the onset of phase separation. In the medium curvature regime we showed computationally and predicted analytically the promotion and inhibition of phase separation based on a geometric lensing effect. We explain this geometric lensing on the sphere and gyroid surface by the transition from Euclidean geometry to spherical and hyperbolic geometry, respectively. At high curvature we observed two different dominant topological effects. On positive curvature we reproduced the spontaneous microswarming reported in Ref.~\cite{BG2017}, which prevents the formation of dense clusters. On negative curvature the particles are more likely to form clusters based on a change in connectivity. Moreover, we investigated the transition from solid to fluid clusters by introducing curvature. We showed that curving the underlying surface is a mechanism to incorporate geometrically induced defect patterns, which locally melt the active crystal.

Based on these results we managed to design four new non-equilibrium phases. Due to the different onset of clustering in different regions of the surfaces we triggered cyclic phase separation on a spherocylinder, cluster waves on a spherocone, cluster fluctuations on a metaball surface and positional clustering around a sinusoidal peak. In general our findings considerably broaden the understanding of the effect of curvature on the steady states of self-propelled particle systems. In particular, the combination between fluidization and the mutable onset of phase separation suggests curvature being a promising tool to control active particle systems for colloidal robotic applications without requiring highly complex particle interactions \cite{RCN2014, YB2020}. For instance, CPS on the spherocylinder and spherocone implies possible applications in particle transport where SPPs aggregate around a load within a positively curved environment, transport it and release it by dispersing due to negative curvature. Overall curvature navigation can be seen as an intrinsically driven alternative to colloidal swarm control via external stimuli \cite{PSSPC2013, ZGMG2018, YAS2019, YZ2020, CLXLYDH2021}. \\

\section*{Acknowledgements}
The authors thank Corwin Kerr, Sophie Youjung Lee and Gabrielle Jones for helpful discussions.
This work was supported as part of the Center for Bio-Inspired Energy Science, an Energy Frontier Research Center funded by the U.S. Department of Energy, Office of Science, Basic Energy Sciences under Award \# DE-SC0000989.  The work was conceived with support from a grant from the Simons Foundation (256297, SCG).  This work used the Extreme Science and Engineering Discovery Environment (XSEDE), which is supported by National Science Foundation grant number ACI-1548562; XSEDE award DMR 140129. Computational resources and services were also supported by Advanced Research Computing at the University of Michigan, Ann Arbor.\\

\begin{appendices}

\section*{Appendix:  Mathematical Background and Model}
\label{sec:Math}
\noindent In our molecular dynamics simulations we model the self-propelled particles (SPP) as spherical objects with diameter $\sigma$. To study phase separation at different curvature profiles, we confine the positions of all particles  $\mathbf{r}_i = (x,y,z)^T$ to either (I) spheres with radius $R_c$ (positive Gaussian curvature $K=\frac{1}{R_c^2}>0$)
\begin{equation}   
0 = x^2 + y^2 + z^2 - R_c^2,
\end{equation}
(II) gyroid minimal surfaces with unit cell length $a$ and surface frequency $\omega=\frac{2\pi}{a}$ (negative Gaussian curvature $K=-\frac{1}{R_c^2}<0$)
\begin{equation}
0=\sin(\omega x)\cos(\omega y){+}\sin(\omega y)\cos(\omega z){+}\sin(\omega z)\cos(\omega x)
\end{equation}
or (III) the two-dimensional flat plane ($K=0$). In the negative curvature simulations the unit cell length $a=\frac{L}{2}$ is coupled to the simulation box with size $L$ such that it matches the periodicity of the surface and always contains 8 unit cells of the gyroid in a $2{\times}2{\times}2$ arrangement unless specifically mentioned otherwise. As $K$ is not constant throughout the gyroid surface, we define the curvature radius in terms of the mean Gaussian curvature: $R_c =  \sqrt{-\langle K \rangle^{-1}}$. The linear relation between $a$ and $R_c$ is derived from the Gauss-Bonnet theorem $R_c=\sqrt{-\frac{A_0}{2\pi\chi}}a\approx0.3508a$ \cite{A1940}, where $\chi=-2$ is the Euler characteristic and $A_0a^2=3.092124a^2$ is the surface area of the gyroid per unit cell \cite{S-TRCH2003}.\\
For the simulations on the four exemplary surfaces, which feature the four novel phase separated states, we restricted the positions of all particles to a spherocylinder (cyclic phase separation))
\begin{equation}
\label{eq:Spherocylinder}
0= \begin{cases} x^2 + y^2 + (z-\frac{l}{2})^2 - R^2 &\mbox{if } z> \frac{l}{2} \\
x^2 + y^2 + (z+\frac{l}{2})^2 - R^2 &\mbox{if } z < -\frac{l}{2} \\
x^2 + y^2 - R^2 &\mbox{else}
 \end{cases}
\end{equation}
with radius $R$ and length $l$, a spherocone (cluster waves)
\begin{equation}
\label{eq:Spherocone}
0= \begin{cases} x^2 + y^2 + (z-\frac{l}{2})^2 - r^2 &\mbox{if } z > \frac{l}{2}+ r\frac{R-r}{l}\\
x^2 + y^2 + (z+\frac{l}{2})^2 - R^2 &\mbox{if } z < -\frac{l}{2}+ R\frac{R-r}{l}\\
x^2 + y^2 - \left(\frac{(r-R)z + \frac{l}{2}(r+R)}{\sqrt{l^2 - (R-r)^2}}\right)^2 &\mbox{else},
 \end{cases}
\end{equation}
with two differently sized spherical cap with radii $r$ and $R$ and length $l$, a metaball surface (cluster fluctuations)
\begin{equation}
\label{eq:Metaball}
0 = \frac{1}{x^2 + y^2 + (z-\frac{l}{2})^2} +  \frac{1}{x^2 + y^2 + (z+\frac{l}{2})^2} - \frac{1}{R^2}
\end{equation}
with radius $R$ and distance $l$ between the two metaballs and a sinusoidal peak (positional clustering)
\begin{equation}
\label{eq:Peak}
0= \begin{cases} h\cos\left(\frac{2\pi}{l}\sqrt{(x^2+y^2)}\right)^2 - z &\mbox{if } \sqrt{(x^2+y^2)} < \frac{l}{4} \\
z &\mbox{else}
 \end{cases}
\end{equation}
with periodicity length $l$ and height $h$.\\
The equations of motion are described by overdamped active Brownian dynamics confined to the different surfaces
 \begin{equation}
\begin{gathered}     
      \gamma \dot{\mathbf{r}}_i =F_i^{\text{Act}}  + F_i^{\text{Man}} +  \sum_j\mathbf{F}_{ij}^{\text{Ex}}\\
      \dot{\theta}_i = \sqrt{2D_r}\Lambda_{r,i},
\end{gathered}
\end{equation}
\noindent
with $\gamma=1$ being the drag coefficient. The inter-particle force $\mathbf{F}_{ij}^\text{Ex}$ is derived from the purely repulsive Weeks-Chandler-Anderson potential \cite{WCA1971}
\begin{equation}
V^\text{ex}(r_{ij})= \begin{cases}4\epsilon\left( \left(\frac{\sigma}{r_{ij}}\right)^{12} - \left(\frac{\sigma}{r_{ij}}\right)^{6}  \right)-\epsilon &\mbox{if } r_{ij}<r_\text{cut} \\
0 &\mbox{else}
 \end{cases}
\end{equation}
with $r_\text{cut}=2^\frac{1}{6}\sigma$ and energy unit $\epsilon=1$. Note that we determine the distance of the particles as the Euclidean distance and not the geodesic distance based on the surface. The active force $F_i^\text{Act} = v_0\hat{\mathbf{v}}_i(\theta_i)$  is given by a unit vector $\hat{\mathbf{v}}_i(\theta_i)=v_i(\hat{\mathbf{e}}_{t_1}(\mathbf{r}_i)\cos\theta_i + \hat{\mathbf{e}}_{t_2}(\mathbf{r}_i)\sin\theta_i)$ and the active force amplitude $v_0$. The unit vector $\hat{\mathbf{v}}_i$ describes the director of the active driving force in terms of the vectors $\hat{\mathbf{e}}_{t_1}(\mathbf{r}_i)$ and $\hat{\mathbf{e}}_{t_2}(\mathbf{r}_i)$, which span the local tangent plane of the confining surface at point $\mathbf{r}_i$. We indicate the ratio between rotational diffusion and active propulsion by the persistence length $l=\frac{v_0}{D_r}=100\sigma$. The rotational diffusion of the director is described by the rotational diffusion coefficient $D_r$ and modeled as uniformly distributed white noise $\Lambda_{r,i}$ such that  $\langle \Lambda_{r,i} \rangle=0$ and $\langle \Lambda_{r,i}^2 \rangle=1$.
The force $F^\text{Man}=\lambda\hat{\mathbf{n}}({\mathbf{r}}_i)$ points in the normal direction $\hat{\mathbf{n}}({\mathbf{r}}_i)$ of the confining surface to keep the SPPs attached to the manifolds. The magnitude $\lambda$ is calculated by using a RATTLE algorithm \cite{PK2016}.\\
The global packing fraction in a system of $N$ particles is approximated by $\phi =\frac{N\pi(\sigma/2)^2}{A_\text{surf}}$, where $\sigma$ is the diameter of the particle and $A_\text{surf}$ is the area of the surface to which the particles are confined. The surface of the flat 2D plane, the 2-sphere and the gyroid are identified as $A_\text{pl}=L^2$, $A_\text{sph}=4\pi R_\text{c}^2$ and $A_\text{gyr}=8A_0a^2$, respectively. The simulations were initialized by placing $N$ passive particles on the surfaces with low curvature and low density. Afterwards, $K$ was slowly increased until the desired density $\rho \in [0.1,0.7]$ and curvature radius $R_c\in [\sigma,20\sigma]$ was reached. The exact values are indicated in the phase diagrams of Fig.~2 in the main text. In the next step, we activated the active forces and waited for $5\cdot 10^6\delta t$ with the time-steps $\delta t = 5\cdot10^{-4} \tau$ to guarantee a steady state. Lastly, we sampled the simulations for $1\cdot 10^7\delta t$. We performed all our simulations with the HOOMD-blue simulation toolkit (v2.9) \cite{AGG2020} and measured time in units of $\tau=\frac{\sigma}{v_0}$. We analyzed the simulations with freud \cite{RDHSAG2020} and used the signac \cite{ADRG2018} software package for data management.

\end{appendices}

\bibliography{reference.bib} 

\pagebreak
\widetext
\newpage
\onecolumngrid

\begin{center}
\textbf{\large Supplementary information:\\Curvature-Controlled Geometrical Lensing Behavior in Self-Propelled Colloidal Particle Systems}\\[0.5cm]
Philipp W. A. Sch\"onh\"ofer\\
\textit{\small Department of Chemical Engineering, University of Michigan, Ann Arbor, Michigan 48109, USA.}\\[0.5cm]
Sharon C. Glotzer\\
\textit{\small Department of Chemical Engineering, University of Michigan, Ann Arbor, Michigan 48109, USA.}\\
\textit{\small Biointerfaces Institute, University of Michigan, Ann Arbor, Michigan 48109, USA.}\\
\end{center}

\setcounter{equation}{0}
\setcounter{figure}{0}
\setcounter{table}{0}
\setcounter{page}{1}

\title{Supplementary information:\\Curvature Controlled Phase Separation in Active Colloidal Particle Systems}%

\author{Philipp W. A. Sch\"onh\"ofer}%
\affiliation{Department of Chemical Engineering, University of Michigan, Ann Arbor, Michigan 48109, USA.}
\author{Sharon C. Glotzer}%
\affiliation{Department of Chemical Engineering, University of Michigan, Ann Arbor, Michigan 48109, USA. and}
\affiliation{Biointerfaces Institute, University of Michigan, Ann Arbor, Michigan 48109, USA.}
\date{\today}%

\vspace{0.5cm}

\section*{Persistence length}
\begin{figure}[h]
\centering
\includegraphics[width=0.48\textwidth]{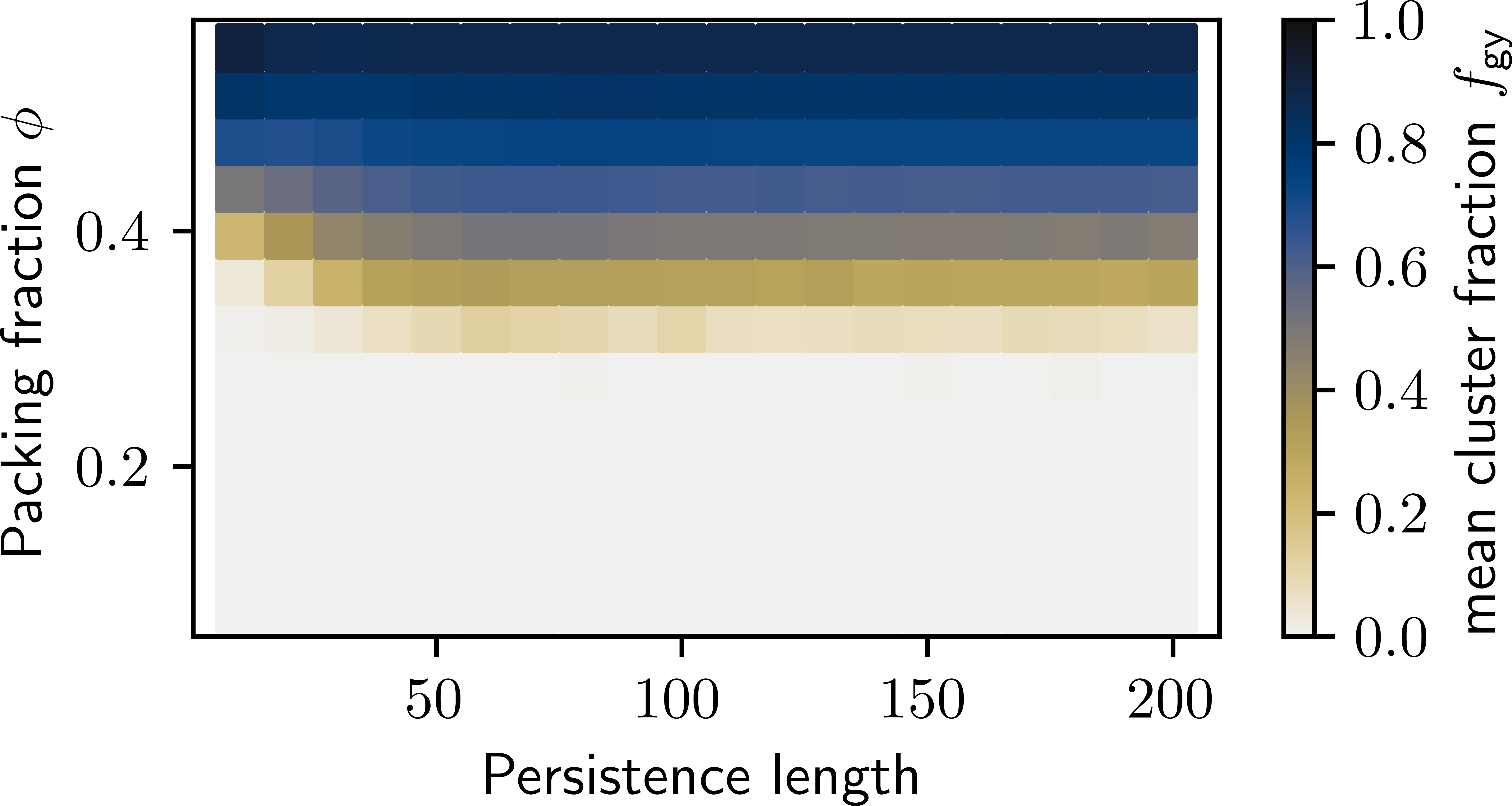}
\includegraphics[width=0.48\textwidth]{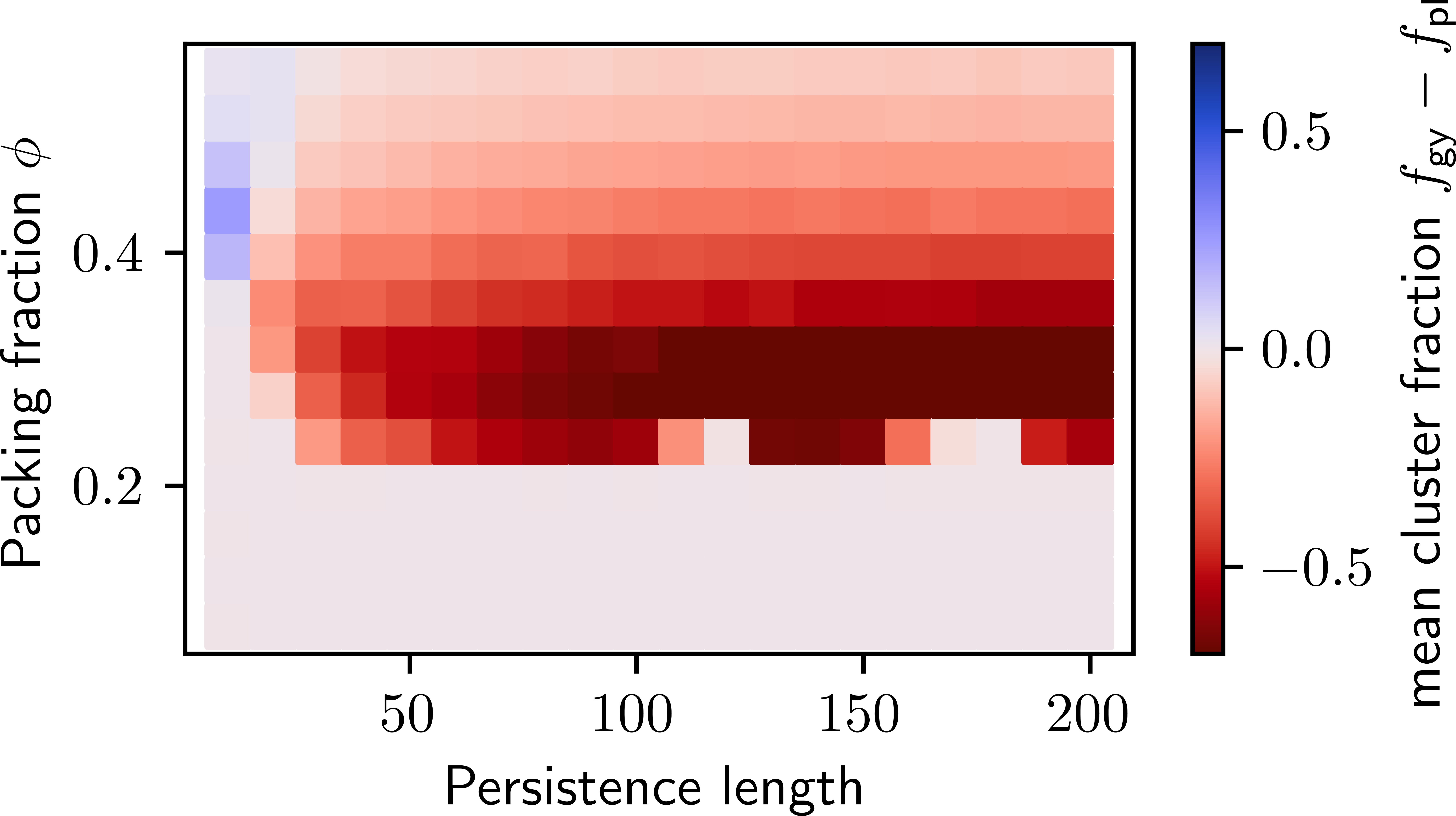}
  \caption{Left: Color map of the mean cluster fraction $f$ for SPPs on a $2\times 2\times 2$ periodic gyroid with mean curvature $K=0.015\sigma^{-2}$. The phase space is sampled based on the packing fraction $\phi$ and the persistence length $l$. Right: Comparison color map between the weighted cluster fraction of the phase diagram on the left and of a corresponding system, where the SPPs are confined to a flat plane with equal surface area as gyroid surface, respectively. Here, blue regions indicate curvature regions with higher mean cluster fraction than on their flat counterpart whereas red domains correspond to lower clustering.}
  \label{fig:Persistance}
\end{figure}

\section*{High curvature regime on the gyroid surface}

\begin{figure}[h!]
\centering
  \includegraphics{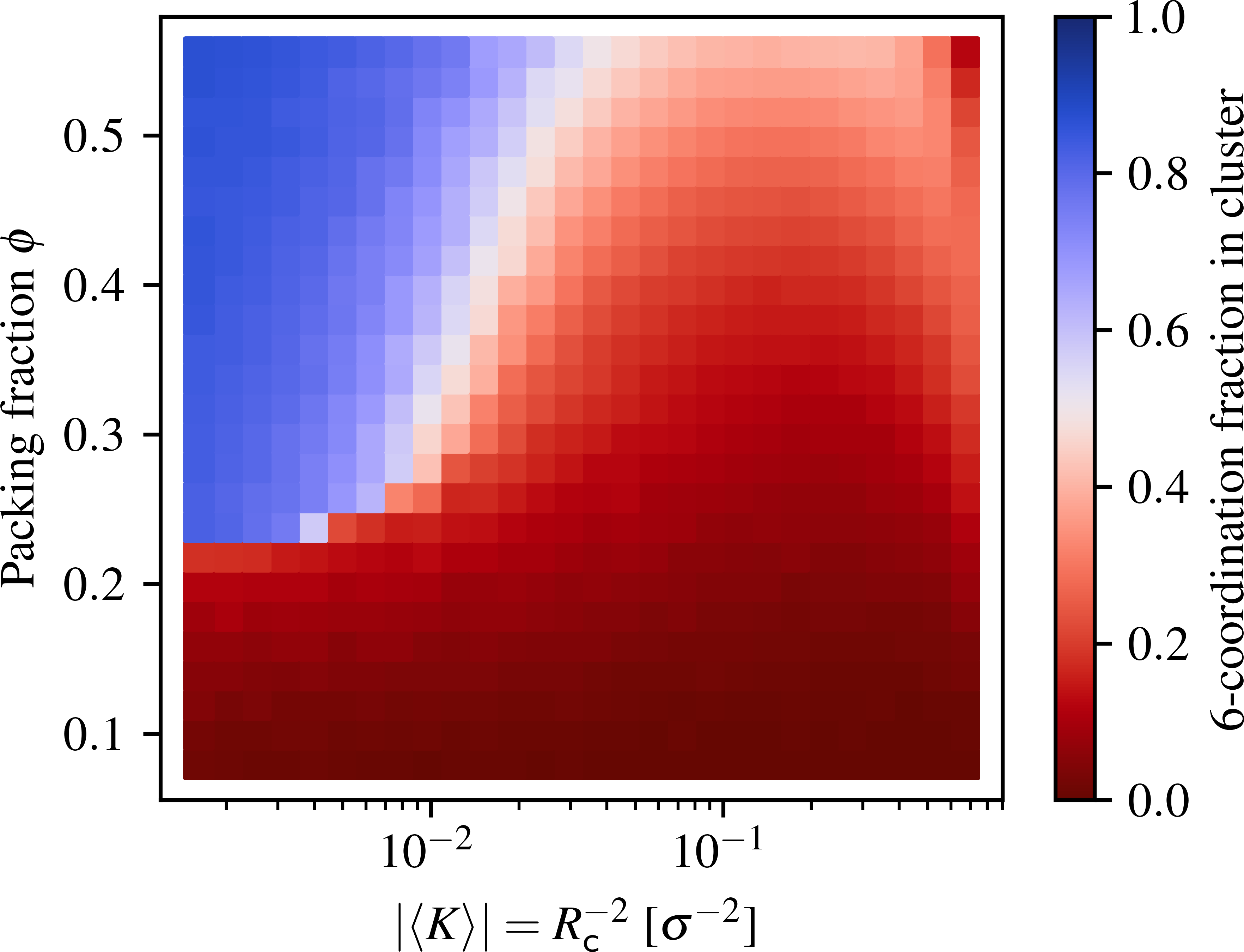}
  \caption{Color map of the weighted mean coordination number for SPPs within the largest cluster on a $2\times 2\times 2$ periodic gyroid surface. The phase space is sampled based on the packing fraction $\phi$ and the mean curvature $|\langle K\rangle |$. In the high curvature regime the SPPs fail to form tightly packed hexagonal configurations, but rather under-coordinated, percolating networks. The arrangements are highly disordered and reminiscent of geometrically frustrated amorphous glasses.}
  \label{fig:Coordination}
\end{figure}

\newpage
\section*{Geometric fluidization}
\label{sec:Fluidisation}

\begin{figure}[h]
\centering
\includegraphics[width=0.49\textwidth]{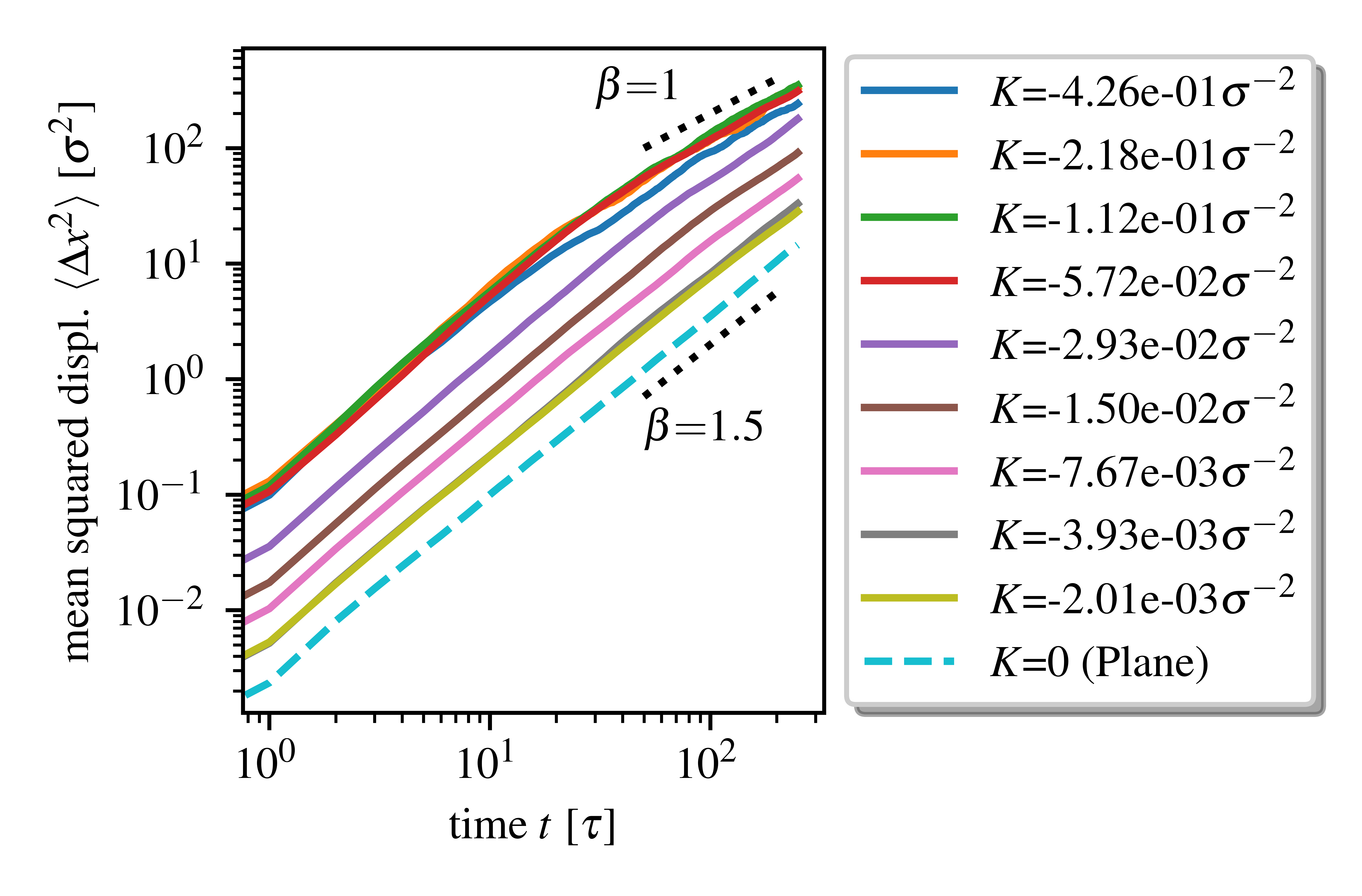}
\includegraphics[width=0.49\textwidth]{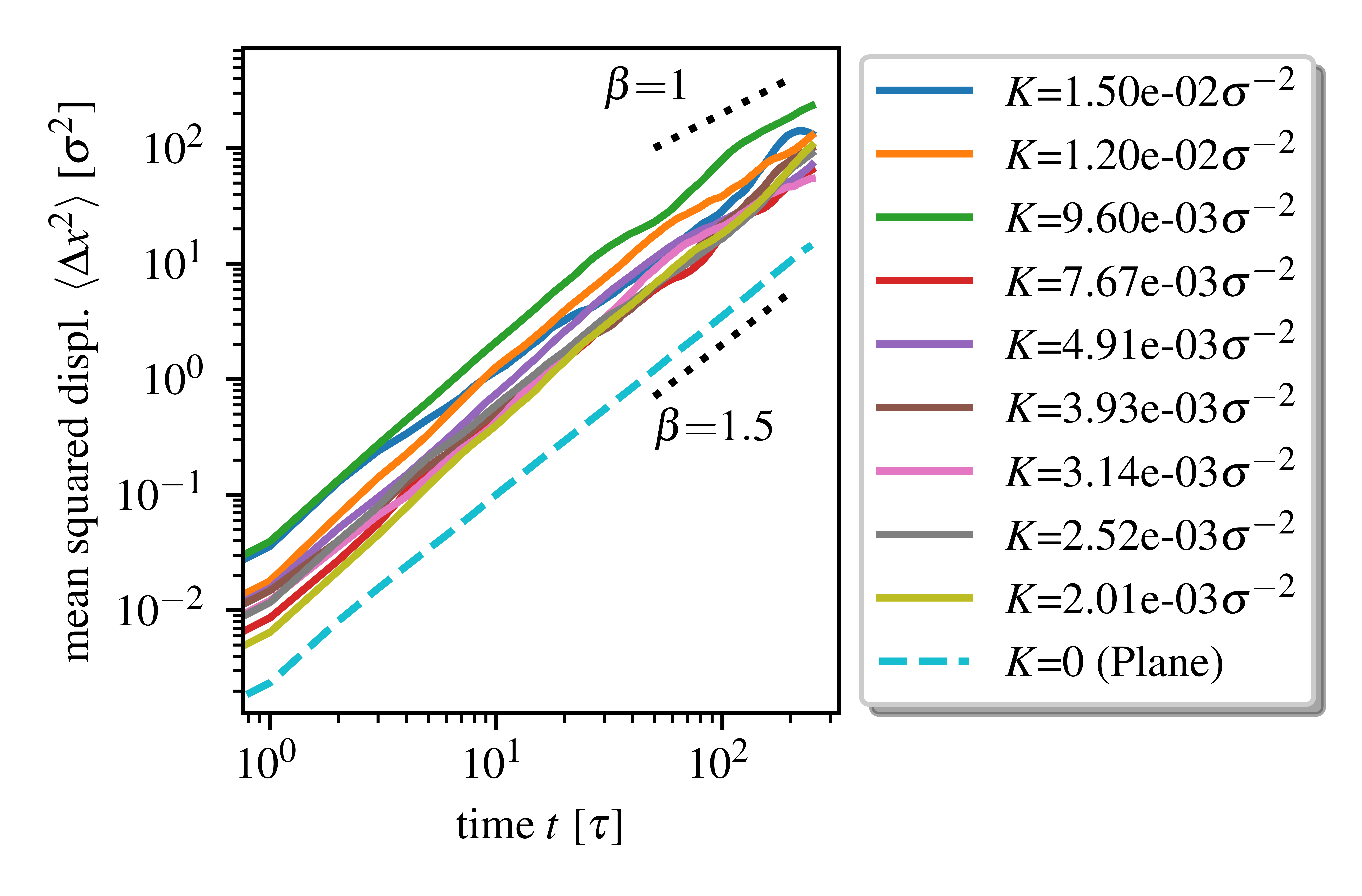}
  \caption{Log-log plot of mean squared displacement of SPPs within the bulk of the dense cluster at packing fraction $\phi=0.5$. The particles live on a $3\times 3\times3$ periodic gyroid surface a) or a sphere b) with varying mean curvature $K$ (solid lines) or on a flat surface (dashed line). The dotted lines indicate a super-diffusive transport of particles at low $K$ and a diffusive behavior at high $K$, where $\beta$ is the exponent of $\langle \Delta x^2\rangle\propto t^\beta$. On a flat two dimensional surface, the phase separated particles are partly situated within a hexagonally ordered lattice structure. Hence, SPPs within the cluster bulk are mostly trapped by their six neighbors causing solid clusters with few internal rearrangements. However, occasionally the interplay between the perpetually changing active forces creates an imbalance within the hexagonal cage, which generates defects. These defects are mostly dislocations that diffuse through the crystalline cluster. The abrupt displacement of particles within the crystal, which is accompanied by dislocation or void defect migration, is the only observed means of propagation of the single particles apart from the collective translation or rotation of the whole cluster. Those L\'evy-flight-like dynamics in flat space are in accordance with the mean squared displacement of the particles within the cluster. On both positively and negatively curved surfaces we observe enhanced motility of the clustered particles. Even though the internal structure of the clusters is still based on hexagonal order, the topology of the sphere and gyroid surfaces prohibits the formation of a perfect hexagonal mono-crystalline domain according to Euler's formula. Instead, the dense cluster must incorporate defect scars. These defect patterns destabilize the crystal in its close vicinity and enhance the mobility of particles as observed in the mean squared displacement. At low and intermediate curvature the scar patterns are distributed such that crystalline regions are interrupted by superdiffusive regions. This behavior is similar to the mechanism of grain boundary melting and can also be observed at the interface of two colliding dense clusters in flat space. Overall the mean squared displacement show that the particles still travel via L\'evy-flight-like transport, yet with increased diffusion coefficient compared to particles confined to a flat surface. In the high curvature regime, the average coordination number within the dense regions in \fig{fig:Coordination} also supports the conclusion that the scar patterns dominate the system and all particles are affected by the geometry induced defects. As hexagonal order becomes less prominent, the system does not separate into a gas and a solid phase but rather into a gas and a high density fluid phase of flowing particles: the fluidizer phase. Fluidization is accompanied by a transition from a superdiffusive L\'evy-flight-like to a diffusive motion.}
  \label{fig:MSD}
\end{figure}

\newpage
\section*{Spectral Analysis}
\label{sec:Spectral }

\begin{figure}[h]
\centering
\includegraphics[width=\textwidth]{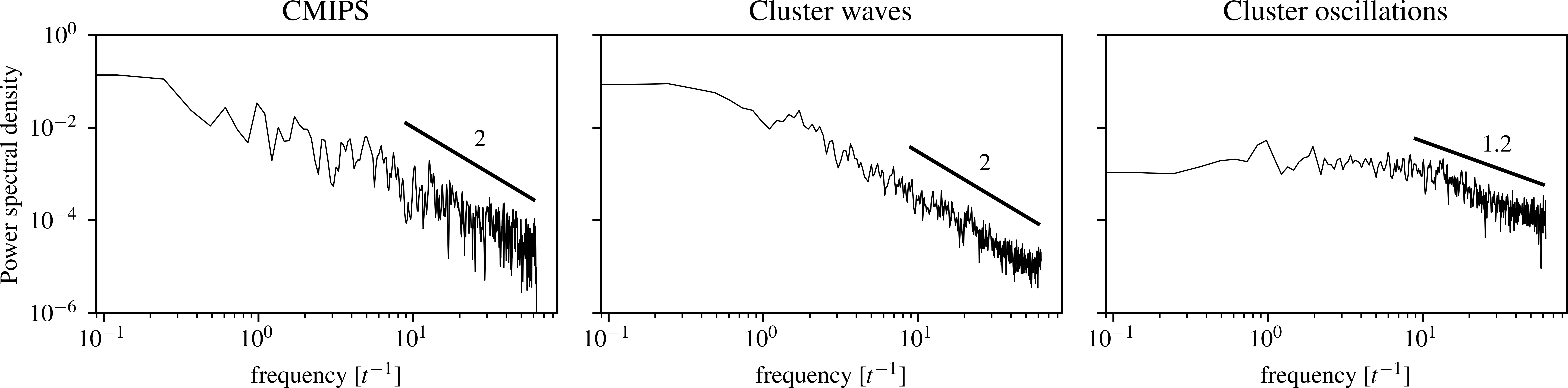}
  \caption{Left: Spectral analysis of the time-resolved cluster fractions of SPPs within the CMIPS, cluster wave and cluster oscillation processes. The corresponding cluster fraction profiles are plotted in Fig. 4 of the main manuscript. The power spectral densities of CMIPS and cluster waves are roughly inversely proportional to the squared frequency indicating red or $\frac{1}{f^2}$-noise. Yet, the cluster oscillations resemble pink or $\frac{1}{f}$-noise more.}
  \label{fig:Spectral}
\end{figure}

\end{document}